# Anomalies in the dynamics of ferrimagnets near the angular momentum compensation point

A.K. Zvezdin[1,2], Z.V. Gareeva [3*], K.A. Zvezdin[1,4]

[1]Prokhorov General Physics Institute, Russian Academy of Sciences, 119991, Moscow, Russia

[2]P.N. Lebedev Physical Institute of the Russian Academy of Sciences, Leninskiy Prospekt 53, 119991, Moscow, Russia

[3]Institute of Molecule and Crystal Physics, Subdivision of the Ufa Federal Research Centre of the Russian Academy of Sciences, 450075, Ufa, Russia

[4]Moscow Institute of Physics and Technology (State University), 141700, Dolgoprudny, Russia

Corresponding authors: gzv@anrb.ru, zvezdin@gmail.com

**Abstract**

In this paper, we elaborate analytical theory of domain wall dynamics close to the angular momentum compensation point based on non-linear dynamic equations derived from the effective Lagrangian of a ferrimagnet. In the framework of the proposed model, we explore dynamic processes in the Walker and post Walker regimes. Analysis of the precession angle and domain wall's velocity oscillations in post Walker regime in a ferrimagnet is performed. We show that although spin oscillations quench the dynamics of domain walls near the Walker breakdown field, a further increase of the driving magnetic field increases domain wall speed and mobility. An anomalous behavior of domain wall dynamic properties near the angular momentum compensation point in ferrimagnets is discussed**.**

## 1. Introduction

Ultrafast spin dynamics in antiferromagnets (AFMs) and ferrimagnets (FiMs) near compensation temperature are receiving much attention owing to the potential applications for spin electron devices [1- 15]. Recent advances in antiferromagnetic spintronics and femtosecond lasers sources operating with ultrafast speeds and ultrashort times open new horizons for future technologies based on high-speed nanoscale elements in memory devices.

An important attributes of spin dynamics in magnets are velocity and mobility of domain walls (DWs). Magnetic DWs, active nanoscale elements, are well – suited for the concept of racetrack memory, logic units and spintronic memristors [7 - 15]. In this sense, a lot of expectations are related with antiferromagnetic materials where DW dynamics is extremely fast. As known [16 - 19], in antiferromagnetic orthoferrites (YFeO$_3$) DW velocity attains 20 km/s and DW mobility is of the order of 0.1 km/s Oe.

Despite the undoubted advantages of antiferromagnetic dynamics, the difficulties associated with the detection of DWs and their manipulation in AFMs force researchers to turn to ferrimagnetic materials. However, the Walker breakdown velocities in FiMs are much lower (an order of magnitude) than those in AFMs.

The exchange enhancement of dynamic parameters and giant DWs velocities are expected in FiMs near the sublattice compensation point $T_{comp}$ [20]. However, there are still no reliable experimental data in this direction. In particular, this can be explained by the fact that the saturation magnetization vanishes $M_s \to 0$ near the compensation point $T \to T_{comp}$, and the corresponding spin torque, which one can call the Zeeman torque, also tends to zero.

Anomalous spin dynamics in FiMs with strong sublatices coupling can occur at two temperatures: $T_{comp}$, sublattice compensation temperature where $M_1=M_2$, and $T_A$, angular momentum compensation temperature where $\frac{M_1}{\gamma_1} = \frac{M_2}{\gamma_2}$, $\gamma_i$ is the gyromagnetic ratio for the specific sublattice, $M_i$ is the sublattice magnetization ($i$=1,2) [5 - 8].

Near the angular momentum compensation point, the situation to study spin dynamics is more favorable. Recent experiments on domain wall dynamics in ferrimagnetic GdFeCo alloy were reported in [1, 2]. Authors of Ref. [1] measured enhancement of DW velocity close to the angular momentum compensation point. The phenomenon was qualitatively explained in terms of collective coordinates and atomistic spin model simulations.

In our work, we elaborate analytical model based on the effective Lagrangian of FiMs [21] to describe dynamics of FiMs near the angular momentum compensation point $T_A$. Proposed approach allows us to explore the structure of a moving DW and its featured dynamic properties: DW speed, mobility and the angle of precession both for steady state motion and post – Walker oscillatory regime.

## 2. Model

To describe dynamics of a ferrimagnet (FiM) close to the angular momentum compensation point we consider FiM with two magnetic sublattices $M_1$, $M_2$ whose magnetization have different temperature dependences. Such situation is realized in intermetallics, the most of rare earth transition metals alloys (e.g. GdFeCo, TbFeCo, CoGd) and weak ferromagnets (e.g. YFeO$_3$, RFeO$_3$) [1, 16 –23]. Approaching the FiMs compensation temperature AFM ordering tends to be established; and antiferromagnetic $\boldsymbol{L} = \boldsymbol{M}_1 - \boldsymbol{M}_2$ and ferromagnetic vectors $\boldsymbol{M} = \boldsymbol{M}_1 + \boldsymbol{M}_2$ are considered as order parameters.

Dynamics of FiM is used to be described in terms of Landau Lifshitz pair equations for each of $\boldsymbol{M}_i$ sublattices [24, 25]. However, to consider domain wall (DW) motion it is more convenient to use Lagrange formalism in the spherical coordinates. The Lagrangian and the dissipation Rayleigh function of two –sublatticed FiM read

$$L = \sum_{i=1}^{2} \frac{M_i}{\gamma_i} (1 - \cos\theta_i) \dot{\varphi}_i - W_i \qquad (1)$$

$$R = \sum_{i=1}^{2} M_i \frac{\alpha_i}{\gamma_i} \left( \dot{\theta}_i^2 + \sin^2\theta_i \dot{\varphi}_i^2 \right) \tag{2}$$

where $M_i$ is the magnitude of the $i$-th sublattice magnetization; $\theta_i, \varphi_i$ are the polar and azimuthal angles characterizing the orientation of the $i$-th sublattice magnetization; $\alpha_i$ and $\gamma_i$ are the $i$-th subllatice damping parameter and gyromagnetic ratio; $W_i$ is the $i$-th subllatice thermodynamic potential.

To characterize the canting of magnetic sublattices we introduce additional variables $\varepsilon, \beta$ defined as $\theta_1 = \theta - \varepsilon$, $\theta_2 = \pi - \theta - \varepsilon$, $\varphi_1 = \varphi + \beta$, $\varphi_2 = \pi + \varphi - \beta$, where $\theta, \varphi$ are the polar and azimuthal angles of antiferromagnetic vector **L** in the spherical coordinate frame with polar axis oriented along the direction of applied magnetic field. In our problem magnetic field is oriented along the normal to a film (see inset in **Figure 1**), uniaxial magnetic anisotropy of a film is supposed to be strong. In the vicinity of compensation temperature ($T_A$) the canting angles $\varepsilon, \beta$ are assumed to be sufficiently small $\varepsilon \ll 1, \beta \ll 1$ that is valid at $H \ll H_{ex}$, $H_{ex}$ ~50 T (TbFeCo, GdFeCo).

The effective Lagrangian of two- sublatticed uniaxial ferrimagnets was suggested and described in details in Ref. [21]. We include into consideration the non – uniform exchange energy, the in-plane magnetic anisotropy and rewrite the effective Lagrangian [21] as follows

$$L_{eff} = \frac{\chi_\perp}{2} \left( \frac{\dot{\theta}}{\gamma_{eff}} \right)^2 + m \left( H - \frac{\dot{\varphi}}{\gamma_{eff}} \right) \cos\theta + \frac{\chi_\perp}{2} \left( H - \frac{\dot{\varphi}}{\gamma_{eff}} \right)^2 \sin^2\theta - \\ K_u \sin^2\theta - K_\perp \sin^2\theta \sin^2\varphi - A \left( \left( \frac{d\theta}{dx} \right)^2 + \sin^2\theta \left( \frac{d\varphi}{dx} \right)^2 \right) \tag{3}$$

where $m = \frac{M_2 - M_1}{2}$, $M = \frac{M_1 + M_2}{2}$; $\chi_\perp = \frac{M}{H_{ex}}$ is the transverse magnetic susceptibility, $H_{ex}$ is the exchange magnetic field acting between sublattices; $K_u, K_\perp$ are the constants of perpendicular and in-plane magnetic anisotropies, here we assume that $H - \frac{\dot{\varphi}}{\gamma_{eff}} \ll K_u$; $A$ is the exchange stiffness constant, $\theta$ and $\varphi$ are the polar and azimuthal angles of the ferromagnetic vector **m**, **H**=(0,0, $H_z$) is magnetic field applied along the "easy magnetization axis" (geometry of a problem is shown in insert to Fig.1). The effective Rayleigh dissipation function is taken in a form [21]

$$R_{eff} = \frac{\alpha_{eff} M}{2\gamma_{eff}}\left(\dot\theta^2 + \sin^2\theta\cdot\dot\varphi^2\right) \qquad (4)$$

where

$$\alpha_{eff} = \bar\alpha\frac{m}{m-m_0},\ \gamma_{eff} = \bar\gamma\frac{m}{m-m_0},\ \bar\alpha = \frac{\alpha_1}{\gamma_1}+\frac{\alpha_2}{\gamma_2},\ \frac{1}{\bar\gamma}=\frac{1}{2}\left(\frac{1}{\gamma_1}+\frac{1}{\gamma_2}\right),\ m_0 = M\frac{\gamma_1-\gamma_2}{\gamma_1+\gamma_2},$$

$$\bar\gamma_{eff} = \bar\gamma\left(1-\frac{mm_0}{M^2}\right)^{-1}$$

where $m_0$ is the magnetization in the compensation point of angular momentum $T_A$.

Let's consider for the beginning DW dynamics $\theta(t)$ in the Walker approach, i.e. assume that $\dot\varphi = 0$, $\frac{d\varphi}{dx}=0$, then the Euler – Lagrange equation for the function $\theta(x, t)$ reads

$$\frac{\chi_\perp}{\gamma_{eff}^2}\ddot\theta - 2A\frac{d^2\theta}{dx^2} + \sin 2\theta\left(K_u - \frac{\chi_\perp}{2}H^2\right) + mH\sin\theta + \frac{\bar\alpha M}{\bar\gamma}\dot\theta = 0 \qquad (5)$$

At the boundary conditions $\theta(-\infty) = 0,\ \theta(+\infty) = \pi$ we find the particular solution of eq. (5)

$$\theta_0 = 2\mathrm{arctg}\left(\exp\left(\frac{x-\dot q t}{\Delta\sqrt{1-\dot q^2/c^2}}\right)\right), \qquad (6)$$

where

$$\frac{\dot q}{\sqrt{1-\frac{\dot q^2}{c^2}}} = \frac{m}{M}\frac{\bar\gamma\Delta}{\bar\alpha}H, \qquad (7)$$

$$\Delta = \sqrt{\frac{A}{K_u - \frac{\chi_\perp}{2}H^2}},\ c = \bar\gamma_{eff}\sqrt{\frac{2A}{\chi_\perp}}$$

where $q$ is the coordinate of DW centre

Eqs. (6), (7) represent the exact solution of eq. (5). One can check it by the direct substitution of eqs. (6), (7) into eq.(5) [16]. Eq. (6) describes the spin distribution inside the $180^0$ domain wall (DW) moving with the velocity $\dot{q}$. Eq. (5), the double sin-Gordon equation with dissipation, and its solution (6) were obtained and used for description and investigation of DW dynamics in the different systems [16, 26, 27].

In the general case $\dot{\varphi} \neq 0$. We solve the system of Euler-Lagrange equations using the perturbation theory for solitons

$$\theta(x,t) = \theta_0(x - \dot{q}t) + \theta_1(x,t) + ...,$$
$$\varphi(x,t) = \varphi(t) + \varphi_1(x,t) + ... \tag{8}$$

where Euler Lagrange equation for $\varphi$ in which we substitute eq. (6) reads

$$\frac{\chi_\perp}{\gamma_{eff}^2}\ddot{\varphi}\sin\theta_0 + \frac{m}{\gamma_{eff}}\dot{\theta}_0 + \frac{\bar{\alpha}M}{\bar{\gamma}}\dot{\varphi}\sin\theta_0 - 2K_\perp \sin\theta_0 \sin\varphi\cos\varphi - \frac{2\chi_\perp}{\gamma_{eff}}\dot{\theta}_0\cos\theta_0\left(H - \frac{\dot{\varphi}}{\gamma_{eff}}\right) = 0 \tag{9}$$

Accounting (8) we rewrite eq. (5) in operator form $\hat{L}\theta_1 = f$ where $\hat{L} = \partial_x^2 - \frac{1}{\Delta^2}\cos 2\theta_0$, function $f$ depends on $\theta_0(x - \dot{q}t)$ and the system parameters. Solvability condition of such equations is determined by the requirement of orthogonality between functions $f$ and $\frac{d\theta_0}{dx}$: $\int_{-\infty}^{+\infty} f \frac{d\theta_0}{dx} dx = 0$ (Fredholm alternative). Implementing the described procedure for eq. (5) and integrating (9) over $x \in (-\infty, \infty)$, we obtain the system of dynamic equations similar to the Slonczewski equations [25] at $\dot{q}/c \ll 1$

$$\frac{\bar{\alpha}M}{\bar{\gamma}\Delta}\dot{q} = m\left(H - \frac{\dot{\varphi}}{\gamma_{eff}}\right)$$
$$\frac{\chi_\perp}{\gamma_{eff}^2}\ddot{\varphi} - \frac{m}{\gamma_{eff}}\frac{\dot{q}}{\Delta} - K_\perp \sin 2\varphi + \frac{\bar{\alpha}M}{\bar{\gamma}}\dot{\varphi} = 0 \tag{10}$$

Note that at $K_\perp = 0$ eqs. (10) give the exact solutions of eqs. (5), (9). As follows from eq. (10) the DW velocity $\dot{q}$ is proportional to the magnetic torques, the precession rate $\dot{\varphi}$ depends on a pressure on DW exerted by driving field, damping and magnetic anisotropy.

### 3. Domain wall dynamics in the steady state (Walker) and post Walker regimes

To understand the characteristic features of DW dynamics close to the angular momentum compensation point we make several steps.

I) We start with the case when in-plane magnetic anisotropy is absent $K_\perp = 0$. Since $\chi_\perp \ll 1$ and $\frac{1}{\gamma_{eff}} \to 0$ close to the angular momentum compensation point $m \to m_0$ one can neglect the first term in eq. (5). In this case in accordance with eqs. (10) DW velocity is determined as in Ref. [1] by equation

$$\frac{\dot{q}}{\Delta} = \frac{m}{M} \bar{\gamma} \frac{\bar{\alpha}}{\bar{\alpha}^2 + \left(\frac{m-m_0}{M}\right)^2} H \qquad (11)$$

The dependences of DW velocity on the parameter $\nu = m/M$ at various magnitudes of magnetic field $H$ are shown in **Figure 1**. For calculations we used the parameters of GdFeCo [1, 21, 22]

$$K_u \sim 1 \cdot 10^5 \frac{erg}{cm^3}, A = 1 \cdot 10^{-6} \frac{erg}{cm}, \chi_\perp \sim 2 \cdot 10^{-3}, M = 1100 G, \bar{\alpha} \sim 0.02, \bar{\gamma} \sim 2 \cdot 10^7, g_d = 2.2, g_f = 2$$

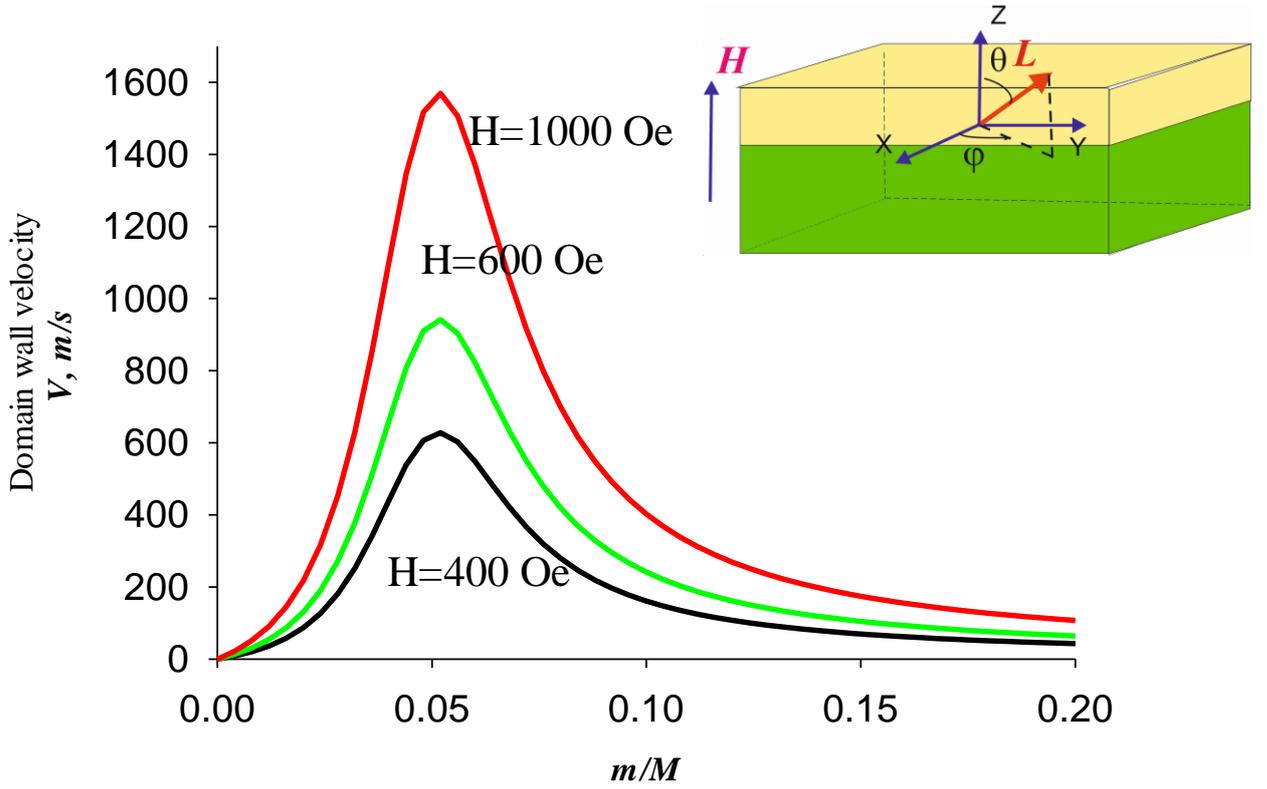

**Figure 1**. Domain wall velocity as the function of the $\frac{m}{M} = \frac{M_1 - M_2}{M_1 + M_2}$ (dimensionless FiM's magnetization) in the case of non – stationary motion at different values of driving magnetic field: black line stands for *H*=400 Oe, green line stands for *H*=600 Oe, red line stands for *H*=1000 Oe. Geometry of the problem is shown in the inset.

As seen in Fig.1 the DW velocity enhances approaching the angular momentum compensation point $v \to v_0$ ($m \to m_0$) and tends to zero at magnetization compensation temperature ($M_1(T_c)=M_2(T_c)$) $v \to 0$ ($m \to 0$). Our results (**Figure 1**) are in consistence with the findings of Ref. [1], the dependences shown in **Figure 1** can be reconstructed in terms of *T* accounting the temperature dependences of the sublattices magnetizations $M_1(T)$, $M_2(T)$.

II) We analyse the dynamic impact given by in-plane magnetic anisotropy $K_\perp \neq 0$.

i) At first we consider the steady state motion occurring in magnetic fields lower than the Walker field $0 < H < H_w$. In this case $\dot{\varphi}_0 = 0$ the DW moves with the velocity $\dot{q} = \mu H$ (see eq.

(10)), where $\mu = \nu \dfrac{\overline{\gamma}\Delta}{\overline{\alpha}}$ is the DW mobility.

Steady – state motion velocity linearly depends on magnetic field up to the Walker velocity

$$V_w = \frac{m}{M} H_w \frac{\overline{\gamma}\Delta}{\overline{\alpha}} \qquad (12)$$

established at the Walker field

$$H_w = \frac{|\alpha_{eff}|}{m^2} K_\perp M = 2\pi |\alpha_{eff}| M \quad . \qquad (13)$$

if we take $K_\perp = 2\pi m^2$

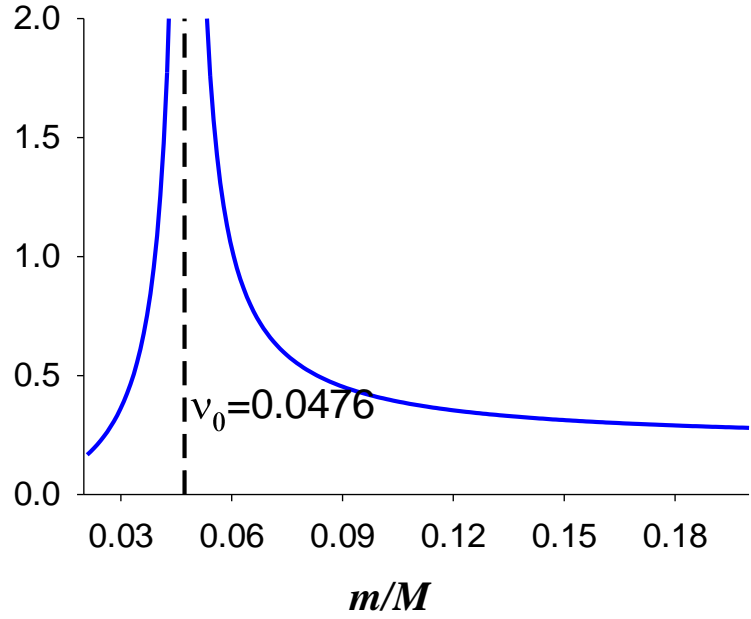

**Figure 2**. The Walker field and the effective mass as functions of $\dfrac{m}{M} = \dfrac{M_1 - M_2}{M_1 + M_2}$ (dimensionless FiM's magnetization).

.

As seen from eq.(13) the Walker field diverges close to the angular momentum compensation point $m \to m_0$ (**Figure 2**).

ii) We turn to the non – stationary regime $\dot{\varphi} \neq 0$ in magnetic fields higher than the Walker field

$H > H_w$. In the case $\chi_\perp \ll 1$ one can neglect the $\frac{\chi_\perp}{\gamma_{eff}^2}\ddot{\varphi}$ term in eqs. (10) and obtain the solution

$$\tan\varphi = \frac{H_w}{H} + \sqrt{1-\left(\frac{H_w}{H}\right)^2}\tan\left(|\gamma_{eff}|Ht\frac{v^2}{v^2+\alpha_{eff}^2}\sqrt{1-\left(\frac{H_w}{H}\right)^2}\right) \qquad (14)$$

$$\frac{\dot{q}}{\Delta} = \gamma_{eff}H_w\frac{v^2}{v^2+\alpha_{eff}^2}\frac{\alpha_{eff}}{v}\left(\frac{H}{H_w}+\left(\frac{v}{\alpha_{eff}}\right)^2\sin 2\varphi\right) \qquad (15)$$

Equations (14), (15) describe the oscillations of precession angle $\varphi$ and DW velocity $\dot{q}$. That means that in post-Walker regime at $H > H_w$ the angle of in-plane precession $\varphi$ and DW velocity $\dot{q}$ become the periodical functions dependent on system parameters. The dependences of oscillations amplitude and period on parameter $v = m/M$ are shown in **Figure 3**. One can see that DW velocity tends to be constant in a time close to the angular momentum compensation point $T_A$. However, due to an unlimited increase in the oscillation frequency approaching $m \to m_0$, a rapid change in the angle of precession at $t \to T$ (green line in **Figure** 3a) manifests itself in the sharpening "beak - like" areas occurring at $t \to T$ on the dependence $\dot{q}(t)$ (green line in **Figure** 3b), where $T$ is the oscillations period

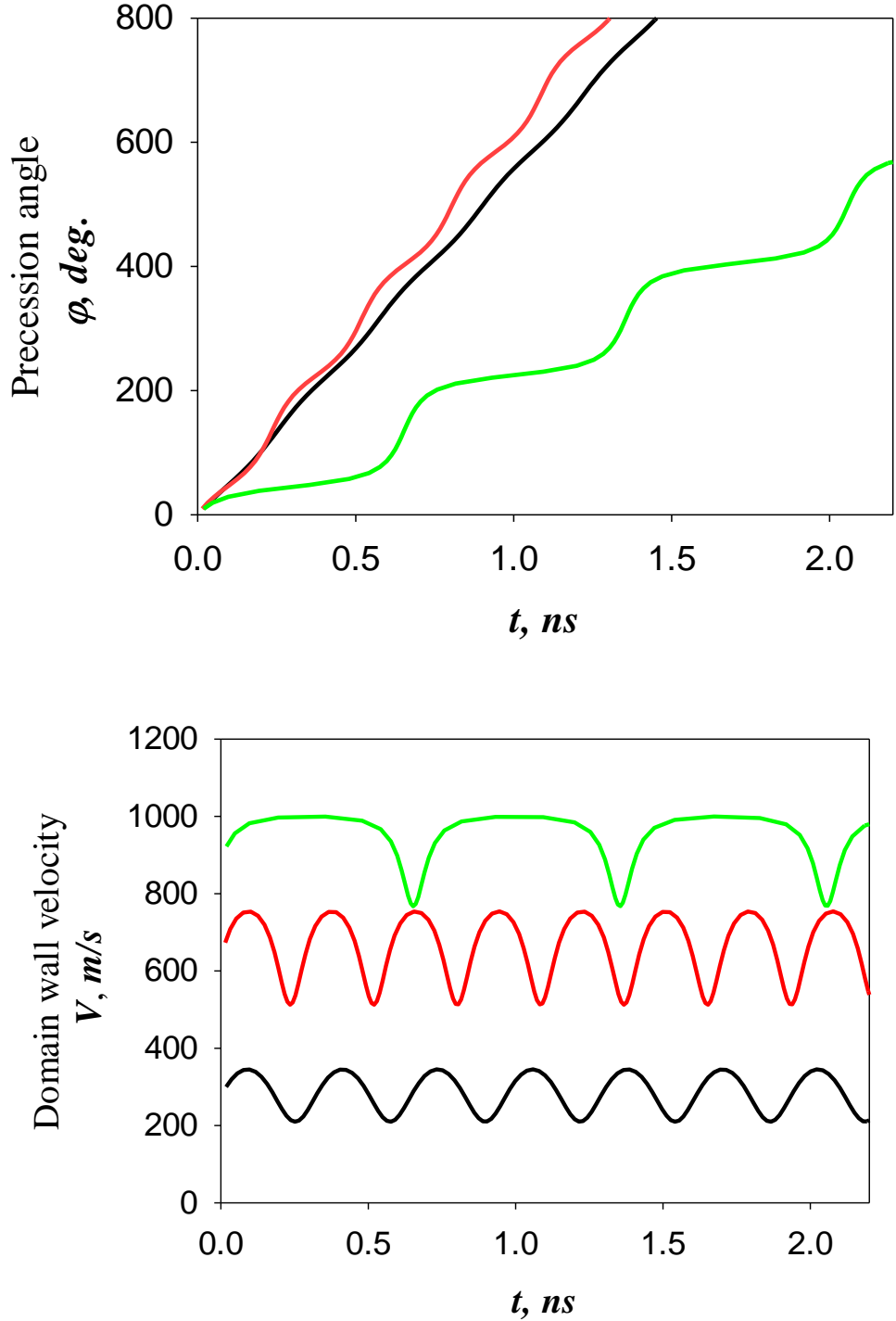

**Figure 3** a) Oscillations of in – plane precession angle as the function of a time, b) domain wall velocities as the functions of a time at different values of the $v = \dfrac{M_1 - M_2}{M_1 + M_2}$: black lines stand for $v = 0.025$, red lines stand for $v = 0.035$, green lines stand for $v = 0.04$, $H$=800 Oe.

Averaging of DW velocity (15) on the precession movement of azimuthal angle $\varphi$ given

by eq. (14) yields

$$\langle \dot{q} \rangle = \gamma_{eff} \Delta H_w \frac{v^2}{v^2 + \alpha_{eff}^2} \frac{\alpha_{eff}}{v} \left( \frac{H}{H_w} + \left( \frac{v}{\alpha_{eff}} \right)^2 \langle \sin 2\varphi \rangle \right) \quad (16)$$

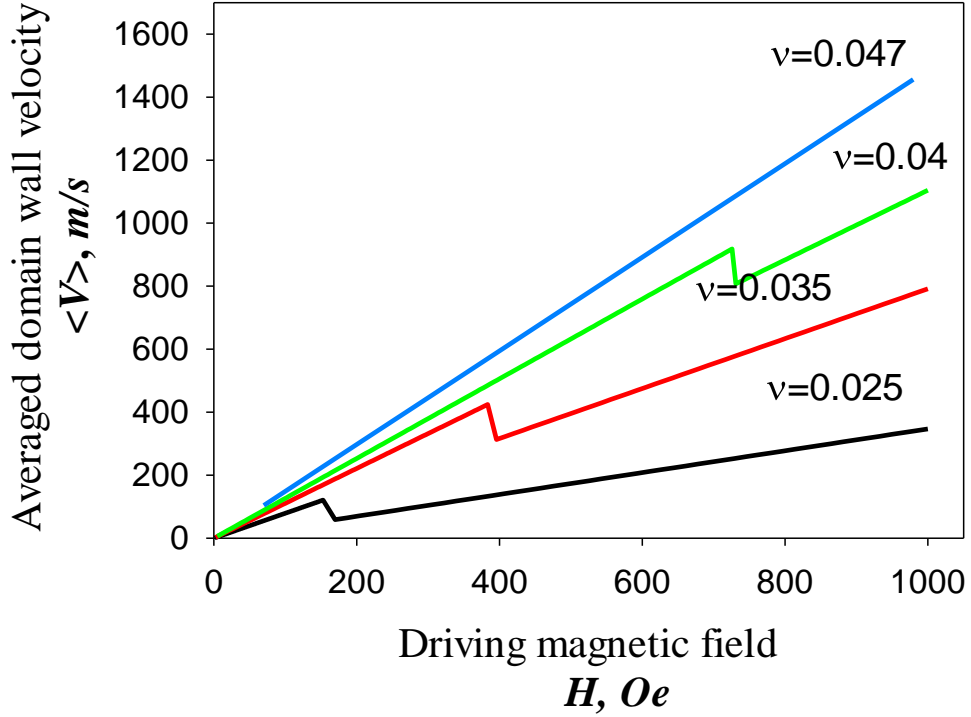

**Figure 4.** Average domain wall velocity as the function of driving magnetic field $H$ in the Walker and post – Walker regimes calculated at different values of the $v = \frac{M_1 - M_2}{M_1 + M_2}$ (dimensionless FiM's magnetization): black line stands for $v = 0.025$, red line stands for $v = 0.035$, green line stands for $v = 0.04$, blue line stands for angular momentum compensation point $v \to v_0 = 0.0475$.

Dependences of the average DW velocity on magnetic field at different $v$ values are shown in **Figure 4** where the Walker and post Walker regimes are distinguished. In the steady state Walker regime DW velocity linearly increases with the driving field $H$ and attains the maximum value $\langle \dot{q} \rangle = V_{max}$ at the Walker field $H = H_w$. In the post Walker regime close to the critical point $H = H_w$ jump – like decrease of DW velocity occurs, however with the enhancement of magnetic field $H > H_w$ average DW velocity $\langle \dot{q} \rangle$ increases. Close to the angular momentum compensation point $\nu \to \nu_0 = 0.0475$ the DW velocity permanently increases.

In contrast to Ref. [1] we show the differences in slope angles of the curves <V>(H) (**Figure 4**) both in Walker and post – Walker regimes and present more detailed consideration of post Walker DW dynamics.

Eq. (7) describes the quasi - relativistic compression of the DW width, which occurs with an increase of the DW velocity. This effect, which is typical for the AFMs dynamics, in particular, leads to saturation of DW velocity with an increase of the driving magnetic field; in AFMs, the DW reaches its ultimate velocity which can be about 20 km/s [16 - 19]. As we showed, in FiMs the situation becomes essentially different: in the vicinity of angular momentum compensation point $m \to m_0$ the Walker field $H_W \to \infty$ and ultimate velocity diverge $c \to \infty$ (see eq. (7)).

## 4 Discussion and conclusion

To conclude, we present the analytical model with integrable dynamic equations on the base of effective ferrimagnet Lagrangian given in Ref. [21]. We obtain the exact solutions of dynamic equations allowing describe DW dynamics in ferrimagnets close to the angular momentum compensation point $T_A$. The processes in-between stationary Walker and non – stationary oscillatory regimes have been explored.

We compare results of our calculations with experimental data on DW dynamics and their explanation given in Ref. [1]. In the case when in-plane magnetic anisotropy is absent the calculated DW velocity and its temperature dependences (**Figure 1**) coincide with the findings of Ref. [1]. We show the principal agreement with results of Ref. [1] on the behavior of DW velocity in the driving magnetic field. Magnetic field applied along the "easy magnetization axis" increases DW velocity up to the Walker breakdown, afterwards the DW quenches due to the transition into oscillatory post – Walker regime typical for ferromagnets. Further enhancement of a field increases

DW velocity. The highest values of velocity (up to 1.5 km per second in a field around 1000 Oe) are attained at the temperatures close to the angular momentum compensation point.

Our analysis also reveals the differences from findings of Ref. [1]. We show that DW velocity is also the function of the net angular momentum $v(T) = (M_2 - M_1)/(M_2 + M_1)$ related with magnetization of specific sublattices dependent on the temperature. In contrast to Ref. [1] we include in our consideration in – plane magnetic anisotropy typical e.g. for uniaxial crystals (GdFeCo) and demonstrate oscillations of the precession azimuthal angle and DW velocity in the post Walker regime. Actually, the in – plane magnetic anisotropy favors outcome of magnetic vector from the DW plane described by azimuthal angle $\varphi$. The competition between impacts given by the driving magnetic field and magnetic anisotropy results in the oscillations of $\varphi$ angle and as a consequence DW velocity close to the Walker breakdown field.

Our findings give additional information on the oscillatory post Walker regime. We show that the amplitude and period of oscillations depend on temperature and material parameters.

Elaborated model can be used as a tool to study plethora of spin dynamics phenomena in ferrimagnetic materials such as magnetic – dipole radiation of electromagnetic waves excited by oscillations of the moving DW, phenomenon theoretically predicted in Ref. [31]; current induced motion of DWs, experimentally observed in AFM coupled bi-layers [2]; accelerating and focusing magnetic vortices and skyrmions [32] attractive for spintronic applications.

**Declaration of Competing Interest**

The authors declare that they have no known competing financial interests or personal relationships that could have appeared to influence the work reported in this paper.

**Acknowledgments**

This work is supported by the Russian Science Foundation grant No.17-12-01333.

# References

[1] K.-J Kim, S.K. Kim, Y. Hirata, S.H. Oh, T. Tono, D.H. Kim, T.Okuno,…, T.Ono, *Nature Materials* **2017**, *16* (12), 1187.

[2] S.A. Siddiqui, J. Han, J.T. Finley, C.A. Ross, L. Liu, *Phys.Rev.Lett*. **2018**, *121*, 057701

[3] B. A. Ivanov, *Low Temp. Phys.* **2019**, *45* (9), 935.

[4] C. D. Stanciu, AV Kimel, F Hansteen, A Tsukamoto, A. Itoh, A. Kirilyuk, Th. Rasing, *Phys. Rev. B* **2006**, *73* (22), 220402.

[5] F. Schlickeiser, U Atxitia, S Wienholdt, D Hinzke, O. Ghubykalo-Fesenko, U. Nowak, *Phys. Rev. B* **2012**, *86* (21), 214416.

[6] M. Binder, A Weber, O Mosendz, G Woltersdorf, M. Izquierdo, I. Neudecker, J. R. Dahn, T. D. Hatchard, J.-U. Thiele, C. H. Back, M. R. Scheinfein, *Phys. Rev. B* **2006**, 74 (13), 134404.

[7] R. F. Soohoo, A. H. J. Morrish, *Appl. Phys.* **1979**, *50* (B3), 1639.

[8] T. Kato, K. Nakazawa, R Komiya, N. Nishizawa, S. Tsunashima, S. Iwata, *IEEE Trans. Magn.* **2008**, *44* (11), 3380.

[9] S. S. Parkin, M. Hayashi, L. Thomas, *Science* **2008**, *320*, 190.

[10] D. A. Allwood, G. Xiong, C.C. Faulkner, D. Atkinson, D. Petit, R.P. Cowburn, *Science* **2005,** *309*, 1688.

[11] J. A. Currivan-Incorvia, S Siddiqui, S Dutta, E.R. Evarts, J. Zhang, D. Bono, C.A. Ross, M.A. Baldo, *Nat. Comm*. **2016,** *7*, 10275.

[12] X. Wang, et al. *IEEE Elec. Devi. Lett*. **2009**, *30*, 294.

[13] J. Munchenberger, G. Reiss, and A. Thomas, J.Appl. Phys. **2012**, *111*, 07D303.

[14] N. Locatelli, V. Cros, J. Grollier, *Nature Materials* **2014**, *13*, 11.

[15] S. Lequeux et al., *Sci. Rep*. **2016**, *6*, 31510.

[16] A.K. Zvezdin. *Pis'ma Zh. Eksp. Teor. Fiz.*, **1979**, *29*, 605, copy in *arXiv:1703.01502* [*cond-mat*] **2017**.

[17] M.V. Chetkin, A.N. Shalygin, *Phys. Sol.State,* **1977**, *19*, 3470

[18] M.V. Chetkin, A. Kampa, *Pis'ma Zh. Eksp. Teor. Fiz.*, **1978**, *27*, 168.

[19] V.G. Baryakhtar, B.A. Ivanov, A.L. Sukstanskii, *Phys. Sol. State* **1978**, *20*, 2177.

[20] B. A. Ivanov, A. L. Sukstanski, *Zh. Eksp. Teor. Fiz.* **1983**, *84*, 370.

[21] M. D. Davydova, K. A. Zvezdin, A. V. Kimel, A. K. Zvezdin, *J. Phys.: Condens. Matter* **2019**, *32* (1), 01LT01.

[22] D.-H. Kim, et al. *Phys. Rev. Lett.* **2019**, *122* (12), 127203

[23] S. K. Kim, *Nature Electronics* **2020**, *3* (1), 18–19.

[24] C. Kittel, Phys. Rev. **1951**, 82, 565

[25] A. P. Malozemoff, J. C. Slonczewski, *Magnetic Domain Walls in Bubble Materials: Advances in Materials and Device Research* (vol.1), Academic Press, **2016**.

[26] A.K. Zvezdin, A.F. Popkov, *Pis'ma Zh. Eksp. Teor. Fiz.* **1985**, *41*, 90.

[27] A.K. Zvezdin, A.A. Mukhin, *Pis'ma Zh. Eksp. Teor. Fiz.* **1985**, *42*, 129.

[28] W. Doering. *Zs. Naturforsch*. **3a**, 373 (1948)

[29] H. E. Khodenkov, *Physica status solidi (a)* **1979**, *53* (1), K103.

[30] X.R. Wang, P. Yan, J.Lu, C.He, *Annals of Physics* 324 (2009) 1815

[31] A.K. Zvezdin, *Pis'ma Zh. Eksp. Teor. Fiz.* **1980**, *31*, 508-510,

[32] S. K. Kim, K.J. Lee, Y. Tserkovnyak, *Phys. Rev. B* **2017,** 95, 140404(R)